 \documentclass[final,5p,times,twocolumn]{elsarticle}

\usepackage{amssymb}
\usepackage{amsmath}

\journal{}

\begin{document}

\begin{frontmatter}



\title{Experimental demonstration of the clock asynchrony model in space-borne gravitational wave detection}

\author[label1]{Ming-Yang Xu}

\author[label1]{Yu-Jie Tan\corref{mycorrespondingauthor}}
\cortext[mycorrespondingauthor]{Corresponding author.}
\ead{yjtan@hust.edu.cn}%

\author[label1]{Ning Ma}

\author[label1]{Ao-Ting Fang}

\author[label1]{Yi-Jun Xia}

\author[label1]{Cheng-Gang Shao}
 \affiliation[label1]{MOE Key Laboratory of Fundamental Physical Quantities Measurement & Hubei Key Laboratory of Gravitation and Quantum Physics, PGMF and School of Physics, Huazhong University of Science and Technology, Wuhan 430074, People's Republic of China}%

\begin{abstract}
Space-borne gravitational wave detection will open the observation window in the 0.1 mHz$-$1 Hz bandwidth, playing a crucial role in the development of cosmology and physics. Precise clock synchronization among satellites is essential for the accurate detection of gravitational wave signals. However, the independent clock counting mechanisms of each satellite pose a significant challenge. This work reports the mathematical model of clock asynchrony, which is mainly dominated by the constant term factor and the linear term factor. Moreover, it experimentally verifies the clock asynchronization technique based on a dual-phasemeter system. Through experimentation, the impacts of these two aspects of clock asynchrony were confirmed, and post-processing techniques were employed to reduce these impacts to as low as $\rm 2\pi \times 10^{-6} rad/Hz^{1/2}@ 3mHz$. Specifically, the constant term factor is measured by Time-delay Interferometry Ranging (TDIR), while the linear term factor can be gauged by clock transmission link. This study provides a reference for understanding the clock asynchrony mechanism and processing clock synchronization issues. Additionally, a low additional noise clock synchronization test system is introduced to support such measurements.
\end{abstract}






\begin{keyword}
Phasemeter;
Clock asynchrony model;
Clock synchronization algorithm;
Space-borne gravitational wave detection.

\end{keyword}

\end{frontmatter}



\section{Introduction}
\label{sec1}

Space-borne gravitational wave detection will open the observation window in the 0.1 mHz$-$1 Hz bandwidth, propelling advancements in cosmology and physics \cite{lisa,tianqin,taiji}. The typical configuration for detecting gravitational waves involves three spacecraft forming an approximately equilateral triangle, utilizing heterodyne laser interferometers to precisely measure the minute displacement changes caused by passing gravitational waves. Each spacecraft is also equipped with an ultra-stable oscillator (USO) to provide an accurate time reference and record the precise moments of data acquisition. Nevertheless, the fact that each spacecraft has its own independent clock counting mechanism presents a formidable obstacle to achieving clock synchronization among different satellites. This challenge is crucial to overcome as accurate clock synchronization is essential for the precise detection of gravitational wave signals \cite{LRR}.

 Pseudo-random code ranging (PRNR) \cite{PRNR} and Time-delay interferometry ranging (TDIR) \cite{TDIR} are currently the two main clock synchronization methods. The former uses a hardware approach to synchronize clocks by comparing the time difference between the transmitted and received pseudo-random codes. The latter uses a post-processing method to synchronize clocks by searching for the lowest noise value of the noise combination at different delays. Experimental verifications related to pseudo-code ranging have demonstrated its excellent ranging performance in weak-light \cite{PRNR1}, bi-directional \cite{PRNR2}, and among three satellites \cite{PRNR3}. Similarly, clock synchronization using the TDIR method has also proven to be a viable and effective approach \cite{TDIR1}.

Naturally, prior to implementing these clock synchronization techniques, it is imperative to determine which specific clock signal is to be synchronized. In the context of space-borne gravitational wave detection, the USO is required to generate multiple frequency components. These encompass the Analog-to-Digital Converter (ADC) frequency utilized for ADC data acquisition, the pilot tone frequency employed to mitigate ADC aperture jitter \cite{pt1,pt2}, and the modulation frequency for incorporating the noise information of the multiplied clock jitter \cite{mod1,mod2}.
Previous research \cite{Unified_model} has indicated that the stability of the pilot tone is the primary source of on-board clock jitter noise. Consequently, the jitter noise of the pilot tone is a clock noise source that demands suppression. This can be achieved through techniques such as the clock noise calibrated time-delay interferometry (TDI) technique \cite{clock_red_1}, the optical frequency comb TDI technique \cite{clock_red_2}, and the sideband arm locking technique \cite{clock_red_3}. The pilot tone correction algorithm can effectively suppress ADC sampling clock noise and aperture jitter noise. As a result, the stability of the ADC clock may seem less important. However, as a marking clock, it is crucial to mark the data and provide a precise timing for each data point.

Space-borne gravitational wave detection missions are based on a constellation of three satellites. Each satellite is equipped with an independent ADC sampling clock, inevitably introducing the issue of clock asynchronization. Clock asynchronization encompasses two key aspects: the constant deviation caused by the inter-satellite delay and the linear deviation resulting from the difference of the central frequencies of the clocks.
Recently, leveraging a hexagonal optical bench \cite{Picometer-Stable-Hexagonal-Optical-Bench}, researchers explored the comprehensive clock asynchronization mechanism among the three satellites \cite{Experimental-verification-of-intersatellite-clock-synchronization-at-LISA}. Based on this, our work further investigates the contributions of both constant and linear terms to clock asynchrony. Theoretically, we derive a coupled noise model that quantifies these terms and establishes synchronization requirements. Experimentally, we verify the impacts of these two terms based on a dual phasemeter system. Emerging research emphasizes the utility of sideband data for transmitting clock information, offering a pathway to enhanced synchronization precision \cite{Ranging_sensor}. To explore this potential, we modulate the pilot tone and the ADC clock under two distinct conditions: homologous (shared source) and non-homologous (independent sources). Our results show that when the ADC clock and the pilot tone are from different sources, one cannot obtain the useful information about clock synchronization, and there is a large amount of clock asynchronization noise in the system. When the ADC clock and the pilot tone are from the same source, one can use the clock comparison information for clock synchronization processing to suppress the clock asynchronization noise in the system. These results can provide a reference for understanding the mechanism of clock asynchrony and the processing of clock synchronization.

\section{The noise model and requirements for clock synchronization}\label{sec2}
\begin{figure}[htbp]
    \centering
    \includegraphics[width=0.75\linewidth]{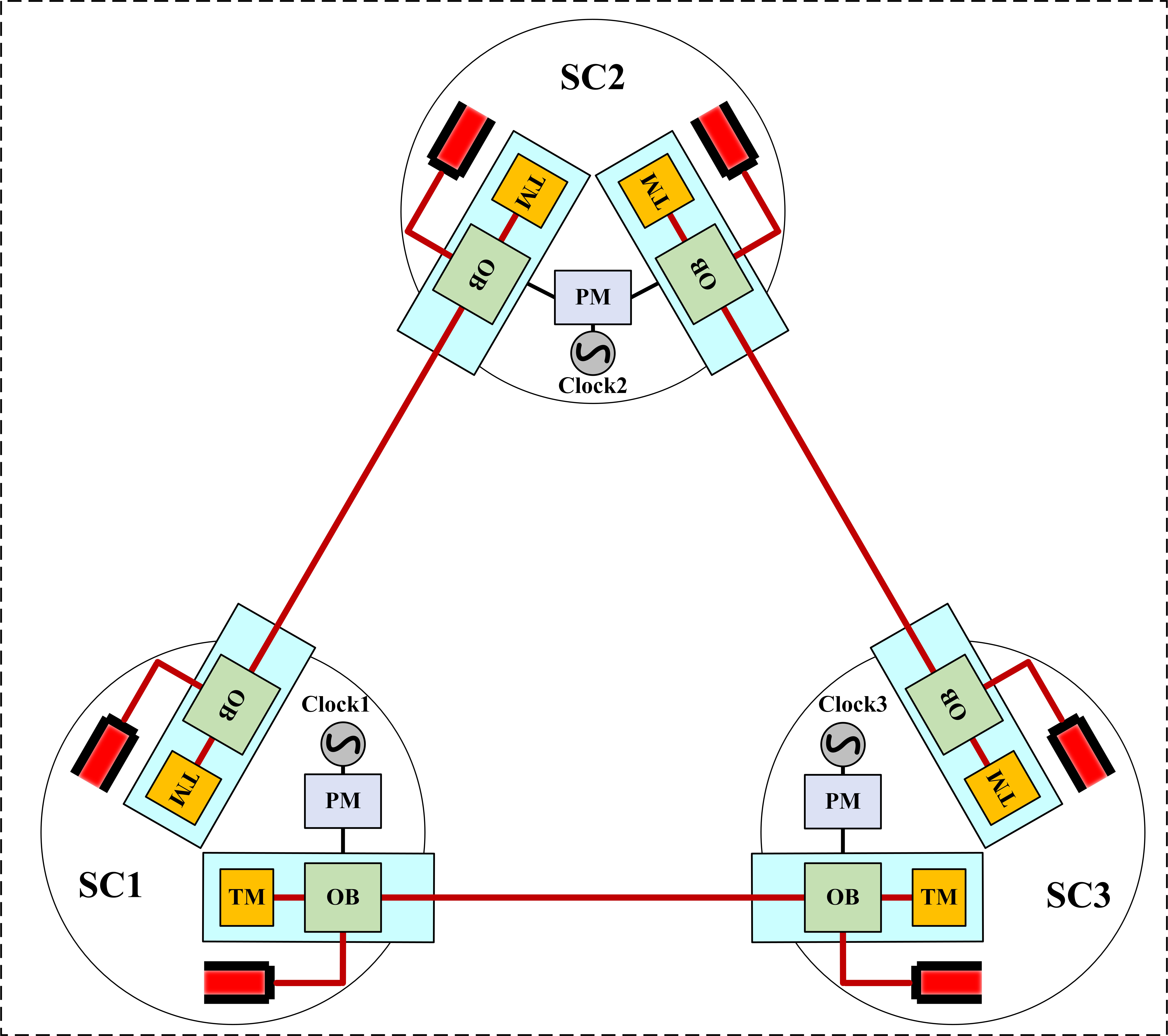}
    \caption{The configuration of space-borne gravitational wave detector. TM: test mass; OB: optical bench; PM: phasemeter; SC: spacecraft.}
    \label{fig:111}
\end{figure}
In space-borne gravitational wave detection, three satellites are deployed, each equipped with an independent USO, as shown in Fig. \ref{fig:111}. Assuming that identical data is collected by each satellite at its respective local time, resulting in the following considerations:
\begin{equation}
\begin{aligned}
f(t_1)=&f(t),\\
f(t_2)=&f(t+t_{2,0}+\alpha_2t+\delta t_2),\\
f(t_3)=&f(t+t_{3,0}+\alpha_3t+\delta t_3),
\end{aligned}
\end{equation}
 where clock 1 serves as the primary clock, $t$ is the clock 1 time, $t_{2,0}$ is the constant initial time offset between clock 1 and clock 2 which primarily arises from the inter-satellite delays and the inherent fixed delay introduced by the disparate hardware components. Additionally, $\alpha_2$ is the linear timing factor of clock 2 relative to the clock 1 which is mainly due to the difference in center frequency between clock 2 and clock 1. Lastly, $\delta t_2$ primarily reflects the stability of clock 2 itself. The parameters for clock 3 are similar to those for clock 2.

\begin{figure}
    \centering
    \includegraphics[width=0.75\linewidth]{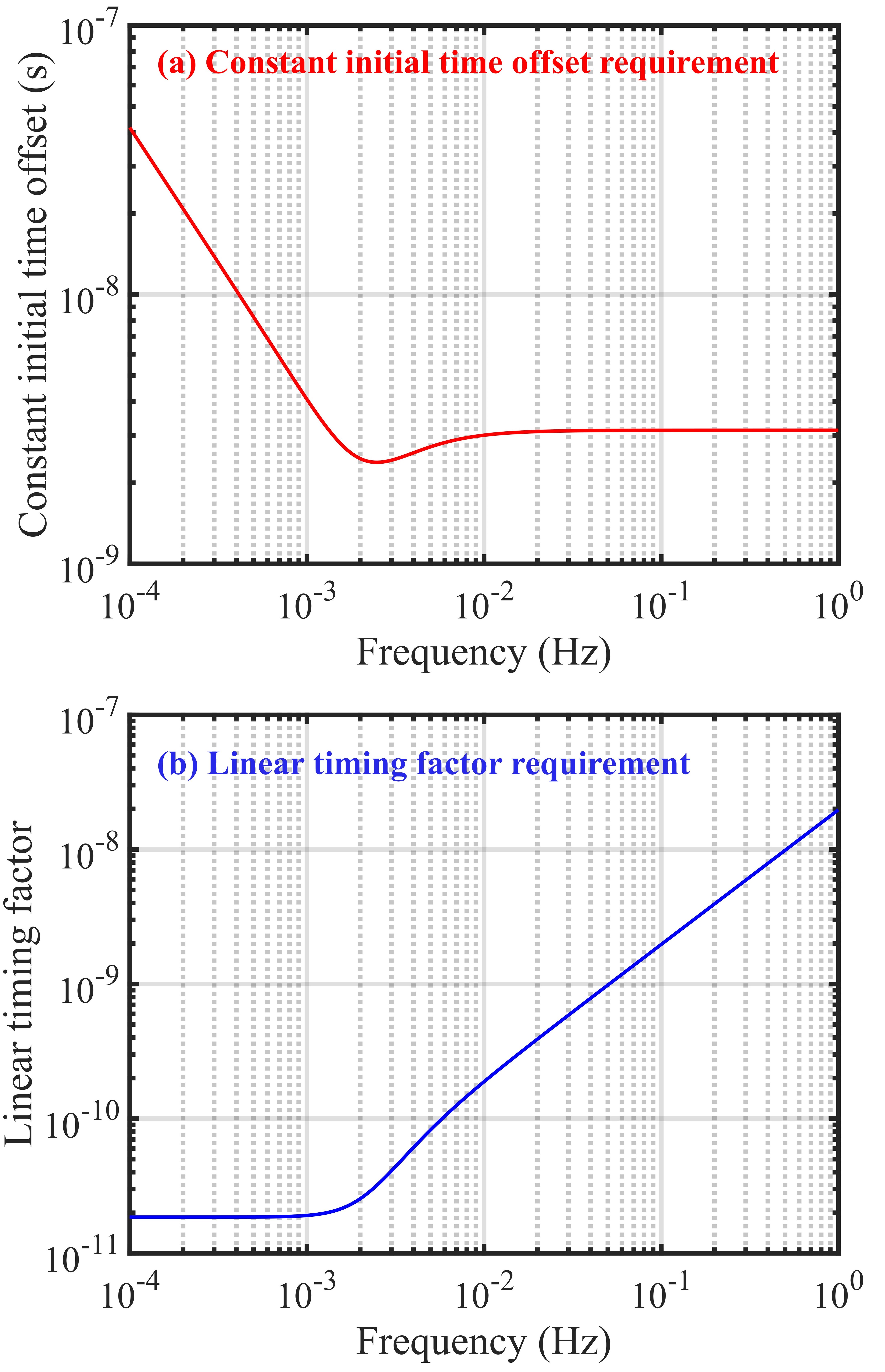}
    \caption{Clock synchronization requirement diagram. (a) The requirement for constant initial time offset; (b) The requirement for linear timing factor}
    \label{fig:1}
\end{figure}

When combining data from two satellites, clock asynchronization inevitably leads to the emergence of residual noise. To delve into this issue, we will focus on the constant and linear terms, while disregarding higher-order terms. The Fourier transform is:
\begin{equation}
\begin{aligned}
f(t)-f(t+t_{2,0}+\alpha_2t)\to F(\omega)-e^{i\omega t_{2,0}}\frac{1}{1+\alpha_2}F(\frac{\omega}{1+\alpha_2})
\end{aligned}
\end{equation}
Then, a rough estimation of the clock synchronization requirements can be undertaken.  In space-borne gravitational wave detection, laser frequency noise is the largest source of noise. The coupling effect between laser frequency noise and the time delay introduced by laser propagation through million kilometers can be reduced by the time delay interferometry (TDI) technique \cite{LRR,wpp}, which will not be considered here. Only the noise coupling effect caused by clock asynchrony is taken into account. Regarding the constant initial time offset requirement, one has:
\begin{equation}\label{eq29}
\begin{aligned}
p(\omega)(1-e^{i\omega t_{0}})&=
p(\omega)(1-{{\rm cos}{\omega t_{0}}}-i{{\rm sin}{\omega t_{0}}})\approx
-i\omega p(\omega)t_{0},
\end{aligned}
\end{equation}
where $p(\omega)$ is the laser frequency noise in frequency domain. The corresponding amplitude spectrum should be smaller than the noise  floor $s_x(\omega)$ of the space-borne gravitational wave detector:
\begin{equation}
\begin{aligned}
t_{0}<\frac{s_x(\omega)}{\omega |p(\omega)|}
\end{aligned}
\end{equation}
Regarding the linear timing factor requirement, one has:
\begin{equation}\label{eq30}
\begin{aligned}
p(\omega)-\frac{1}{1+\alpha}p(\frac{\omega}{1+\alpha})\approx \alpha p(\omega)+\alpha\omega p'(\omega)
\end{aligned}
\end{equation}
where $p'(\omega)$ is the derivative of the Fourier transform $p(\omega)$ of the laser frequency noise with respect to the frequency $\omega$. The corresponding amplitude spectrum should be smaller than the noise floor $s_x(\omega)$ of the space-borne gravitational wave detector:
\begin{equation}
\begin{aligned}
\alpha<\sqrt{\frac{s_x(\omega)^2}{|p(\omega)|^2+\omega^2 |p'(\omega)|^2+2 \omega \text{Re}[p(\omega)p'(\omega)*]}}.
\end{aligned}
\end{equation}
Based on the parameter analysis in typical space-borne gravitational wave detection, the first term in the denominator of the above equation dominates. By taking the pre-stabilized laser frequency noise\cite{laser_requir} and the gravitational wave detector noise floor \cite{noise_requir} parameters as $|p(\omega)|=300\sqrt{1+(\frac{3\rm mHz}{f})^2}\ \rm Hz/Hz^{1/2}$  and $|s_x(\omega)|=\frac{2\pi f}{\lambda}\ \sqrt{1+(\frac{2\rm mHz}{f})^4}\ 1\  \rm pm/Hz^{1/2}$ , respectively, one can obtain a rough estimate of the clock synchronization requirements as shown in Fig. \ref{fig:1} with $t_{2,0}<2.3\ \rm ns$ and $\alpha_2<1.8\times10^{-11}$.

\section{Experiment setup and results for clock synchronization with dual phasemeter}\label{sec3}
\begin{figure*}[htbp]
    \centering
    \includegraphics[width=0.75\linewidth]{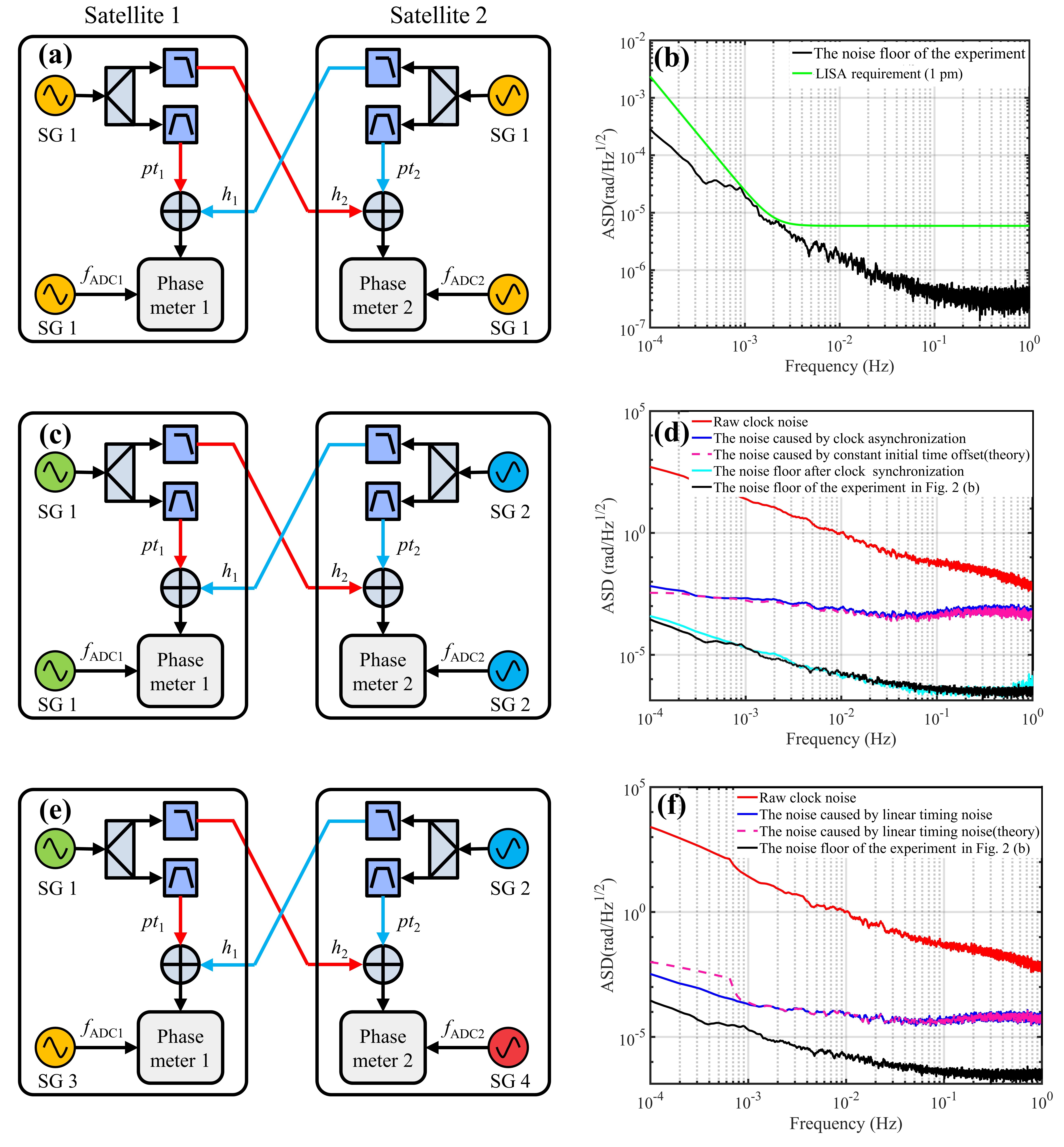}
    \caption{Diagram of experimental test of inter-satellite clock synchronization. (a) Schematic diagram of the experimental noise floor test, where all clock signals share the same source; (b) Experiment results diagram of Fig. \ref{fig:9}(a); (c) Schematic diagram of the experiment where the inner-satellite clocks are synchronized from a common source, whereas the inter-satellite clocks are not synchronized among themselves; (d) Experimental result diagram of Fig. \ref{fig:9}(c); (e) Schematic diagram of the experiment where the ADC clock and pilot tone within the inter-satellite are derived from separate sources, and the clocks between inter-satellite are also independently sourced; (f) Experimental result diagram of Fig. \ref{fig:9}(e).}
    \label{fig:9}
\end{figure*}

The ADC clock, as a marking clock, will record the time of each data point and play a leading role in the clock asynchronization between the two satellites. In this section, a dual phasemeter system is utilized to test clock synchronization issue between two satellites. The setup of the experiment is shown in Fig. \ref{fig:9}.

In this experiment, four signal sources (Rigoal DG4202) and two homemade phasemeters \cite{PM1} are used. Specifically, two of the signal sources are configured to output a 7.667 MHz square wave, each equipped with filters to isolate the 7.667 MHz signal for measurement and to extract the 23.001 MHz harmonic as the pilot tone. The remaining two signal sources contribute 50 MHz outputs, designate as the ADC trigger clocks for the respective phasemeters, effectively simulating the asynchronous clock behavior that can occur between two satellites. Taking Fig. \ref{fig:9}(e) as a typical experimental configuration, the data stream collected by phasemeters comprise:
\begin{align}
    h_1(t_3)=&{\rm7.667MHz}[q_2(t_3)-q_{3}(t_3)+q_{\rm ADC 1}],\\
     pt_1(t_3)=&{\rm23.001MHz}[q_1(t_3)-q_{3}(t_3)+q_{\rm ADC 1}],\\
      h_2(t_4)=&{\rm7.667MHz}[q_1(t_4)-q_{4}(t_4)+q_{\rm ADC 2}],\\
     pt_2(t_4)=&{\rm23.001MHz}[q_2(t_4)-q_{4}(t_4)+q_{\rm ADC 2}],
\end{align}
Among them, $h_1(t_3)$ is the signal generated by signal source (SG) 2 and is sampled by the ADC clock driven by SG 3. Therefore, there will be the clock jitter noise of SG 2, quantified by the relative dimensionless quantity $q_2(t_3)$; the ADC clock jitter noise of SG 3, quantified by the relative dimensionless quantity $q_3(t_3)$, and the aperture jitter noise $q_{\rm ADC 1}$ of ADC1. $pt_1(t_3)$ is the signal generated by SG 1 and serves as the pilot tone. The information contained in the data of $h_2(t_4)$ and $pt_2(t_4)$ is similar to that of $h_1(t_3)$ and $pt_1(t_3)$. After pilot tone correction process, the ADC clock jitter noise and aperture jitter noise can be suppressed:
\begin{align}
   { \rm cor}_1(t_3)&\!=\!h_1(t_3)\!-\!pt_1(t_3)/3\!=\!{\rm7.667MHz}[q_2(t_3)\!-\!q_1(t_3)],\\
    {\rm cor}_2(t_4)&\!=\!h_2(t_4)\!-\!pt_2(t_4)/3\!=\!{\rm7.667MHz}[q_1(t_4)\!-\!q_2(t_4)].
\end{align}

\textbf{Testing the experiment noise floor.} Firstly, the four signal sources are homologous from a same clock as shown in Fig. \ref{fig:9}(a), so that the clock jitter noise and clock asynchronization noise can be common suppressed:
\begin{align}\label{eq37}
     { \rm cor}_1(t)\!+\! { \rm cor}_2(t)\!=\!{\rm7.667MHz}[q_2(t)\!-\!q_1(t)\!+\!q_1(t)\!-\!q_2(t)].
\end{align}
The experiment results are shown in Fig. \ref{fig:9}(b). The black line represents the noise floor of Eq. (\ref{eq37}). The residual noise, primarily originating from the filters, is below the level of $2\pi \times 10^{-6}$ rad/ $\rm Hz^{1/2}$ at 3 mHz. This performance exceeds the requirements for space-borne gravitational wave detection, as indicated by the green line.

\textbf{Conducting clock synchronization via the clock transfer link.} Subsequently, within the same satellite, the pilot tone and the ADC clock are synchronous, whereas the clocks in different satellites are of different origins and lack homology, as depicted in Fig. \ref{fig:9}(c). In this case, one can obtain:
\begin{align}\label{eq38}
    { \rm cor}_1(t_1)\!\!+\!\!{ \rm cor}_2(t_2)\!\!=\!\!{\rm7.667MHz}[q_2(t_1)\!\!-\!\!q_1(t_1)\!\!+\!\!q_1(t_2)\!\!-\!\!q_2(t_2)],
\end{align}
which is dominated by the residual clock noise due to the clock unsynchronization ($t_1 \neq t_2$). As discussed in Ref. \cite{Ranging_sensor}, in space-borne gravitational wave detection, the USO-calibration TDI technique is employed to mitigate significant clock noise. This technique involves modulating and amplifying the clock noise, then transmitting it to the opposing satellite for comparison with the amplified clock noise there. It requires that the modulation frequency and the pilot tone be homologous. To ensure that the ADC frequency and the modulation frequency are derived from the same source, the clock comparison information of the ADC in different satellites can be also obtained via the clock comparison link. This significantly enhances the accuracy of clock synchronization. In this experiment, we ensured that the ADC frequency and the pilot tone originated from the same source and established a clock comparison link. Consequently, the clock comparison information of the ADC can be acquired, which is crucial for clock synchronization processing. The specific process can be found in Ref. \cite{Experimental_demonstration_of_sub_100_picometer} and \ref{scB}, which involves synchronizing the linear timing factor through the clock noise transmission link and the constant initial offset differences through the TDIR algorithm. The experiment results are shown in Fig. \ref{fig:9}(d). The red line represents the raw clock noise. The blue line shows the information of Eq. (\ref{eq38}), while the pink dotted line indicates the theoretical result of the amplitude spectrum of Eq. (\ref{eq29}) (with $t_{2,0}\approx 0.010859 s$). Since the blue line is almost coincident with the pink dotted line, it is evident that Eq. (\ref{eq38}) is primarily dominated by clock unsynchronization noise, which is mainly caused by the constant initial time offset. The light blue line represents the result after clock synchronization, which can reach the noise floor of the experiment, as shown by the black line.

\textbf{Testing the clock asynchrony mechanism.} Finally, the four signal sources in satellite 1 and satellite 2 are independent as shown in Fig. \ref{fig:9}(e):
\begin{align}\label{eq39}
    { \rm cor}_1(t_3)\!\!+\!\!{ \rm cor}_2(t_4)\!\!=\!\!{\rm7.667MHz}[q_2(t_3)\!\!-\!\!q_1(t_3)\!\!+\!\!q_1(t_4)\!\!-\!\!q_2(t_4)].
\end{align}
Since the ADC clock noise transmission link has not been established, the frequency jitter between ADC 1 and ADC 2 cannot be obtained, which prevents us from synchronizing the linear timing factor of the clocks. However, the differences in the constant initial offset of the clocks can still be obtained through the TDIR algorithm. The experiment results are shown in Fig. \ref{fig:9}(f). The red line is raw clock noise, the blue line shows the result of synchronization the differences in the constant initial offset (with $t_{2,0}\approx 0.05452 s$), which is dominated by the linear timing factor unsynchronization noise, and the pink dolt line shows the theoretical result of the amplitude spectrum of Eq. (\ref{eq30}), which is higher than the experiment noise floor as shown in black line. The linear timing factor between the two ADC clocks is measured by another phasemeter, which is not shown in the Fig. \ref{fig:9}(e) and the factor measured in this experiment is about $-1.7918\times10^{-7}$.

\begin{figure*}
    \centering
    \includegraphics[width=0.7\linewidth]{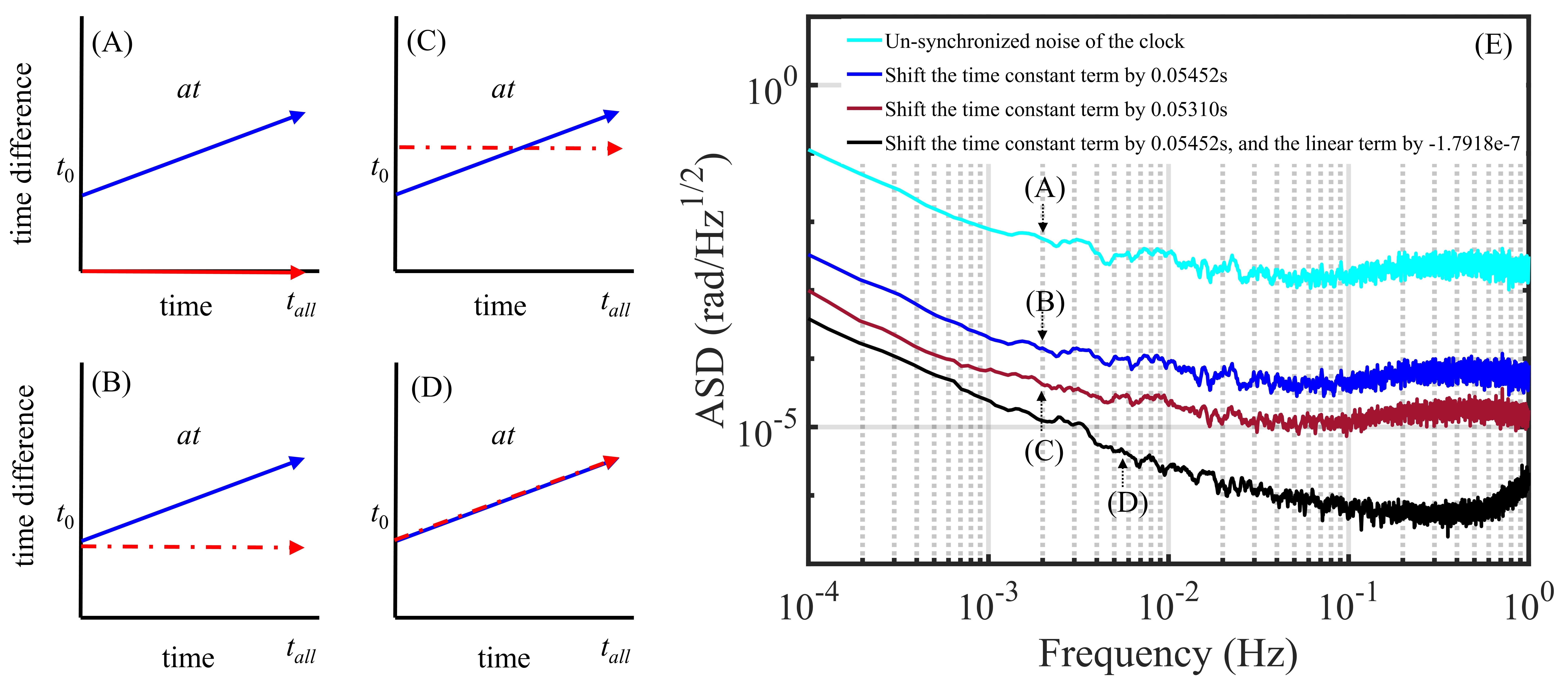}
    \caption{Diagram of verification of the coupling mechanism between the constant term and the linear term. (A) There are both constant term $t_0$ and linear term $at$ deviations of clock 2 relative to clock 1; (B) Compensated for the constant term with $t_0$; (C) Compensated for the constant term with $t_0+\frac{1}{2}\times a\times t_{all}$; (D) Compensated for the constant term with $t_0$ and linear term with $at$; (E) The amplitude spectral density of the test results.}
    \label{fig:10}
\end{figure*}
 To investigate the coupling mechanism under clock asynchrony, a systematic analysis is performed using the configuration in Fig. \ref{fig:9}(e), with results summarized in Fig. \ref{fig:10}.
In Fig. \ref{fig:10}(A), taking $f_{\rm ADC1}$ as the reference clock, the time output profile of which is represented by the red line. In contrast, the time output of the $f_{\rm ADC2}$  clock shows both a constant deviation and a linear deviation compared to that of the $f_{\rm ADC1}$ clock, as illustrated by the blue line, and can be described as follows:
\begin{align}
    t_{\rm ADC2}=t_{\rm ADC1}+t_0+at.
\end{align}
In this case, the uncorrected noise level of Eq. (\ref{eq39}) corresponds to the light blue line in Fig. \ref{fig:10}(E). By compensating a constant time offset of 0.05452 s, the noise level is reduced to the level represented by the blue line in Fig. \ref{fig:10}(E), which matches the blue line in Fig. \ref{fig:9}(f). Additionally, the time output profiles of the two ADC clocks are characterized as shown in Fig. \ref{fig:10}(B).

Further analysis revealed that optimizing the compensation of constant term to 0.05310 s (brown curve in Fig. \ref{fig:10}(E)) achieved additional noise suppression. This value is in agreement with the theoretical expression $t_0+\frac{1}{2}\times a\times t_{all}$, where $t_{all}=15550s$ is the total time and $a=-1.7918\times10^{-7}$. Substituting these parameters yields:
\begin{align}
    0.05452s+\frac{1}{2}\times(-1.7918\times10^{-7})\times15550s\approx0.05310s.
\end{align}
confirming agreement with theoretical predictions (Fig. \ref{fig:10}(C)). Finally, linear term compensation minimized the inter-clock variance (Fig. \ref{fig:10}(D)), effectively suppressing asynchronous noise to the test noise floor (black curve in Fig. \ref{fig:10}(E)).

\section{Conclusion}\label{sec4}
This research addresses the critical issue of clock synchronization in space-borne gravitational wave detection. Initially, it theoretically evaluates the synchronization requirements for the ADC marking clock between satellites, focusing on parameters such as the initial time offset and the linear timing factor. Subsequently, ground experiments are conducted to validate the mechanism of inter-satellite clock synchronization. An experimental setup comprising four signal sources and two phasemeters is designed to simulate clock asynchrony between satellites. The results indicate that the asynchrony of the ADC marking clock is closely related to the constant initial time offset and the linear timing factor. To achieve synchronization, the initial time offset can be determined using the TDIR method, while the linear timing factor can be obtained from clock comparison data. In this study, the linear timing factor between two ADC clocks is measured using an additional phasemeter. In actual space-borne gravitational wave detection, a clock signal comparison link will be established between satellites to provide modulation noise comparison information. If the modulation frequency is synchronized with the pilot tone, and the pilot tone is synchronized with the ADC clock, the comparison information of the ADC marking clocks between two satellites can be directly obtained, which can be used to solve the linear synchronization issue. Thus, aligning the ADC clock with the pilot tone is advantageous for achieving high-precision clock synchronization. Additionally, inter-satellite pseudo-range measurements can also be used to address both the constant time offset and linear time deviations. Therefore, some researchers have explored combining pseudo-range measurements with clock signal comparison information to achieve better synchronization results. The findings of this study are significant for understanding the clock asynchrony mechanism and optimizing synchronization processing, providing valuable insights for the development of space-borne gravitational wave detection.

\appendix
\section{Clock synchronization processing of Fig. \ref{fig:9}(d)}\label{scB}
\begin{figure}
    \centering
    \includegraphics[width=0.75\linewidth]{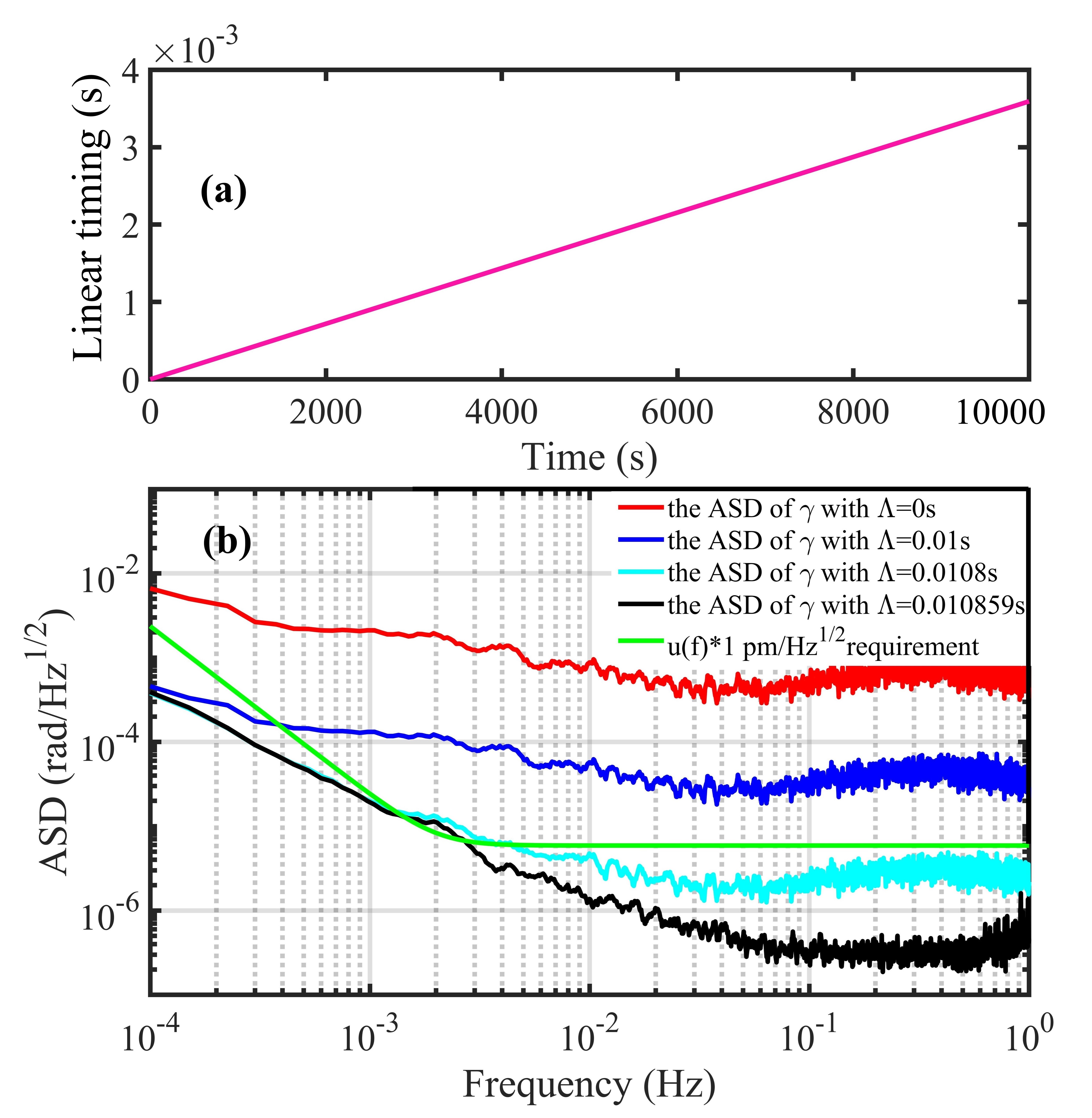}
    \caption{The linear timing factor and constant initial time offset}
    \label{fig:11}
\end{figure}

The difference of linear timing noise between two clocks can be obtained from the data $\rm cor_2$ collected by phasemeter 2, which is:
\begin{equation}
\alpha t \approx\int_{0}^{\tau} \frac{\rm cor_2}{\rm 7.667 MHz} d\tau. 
\end{equation}
The result is shown in Fig. \ref{fig:11}(a). The constant initial time offset between two clocks can be determined by TDIR. Briefly, by minimizing the ASD of the residual noise related to the time-delay parameters, accurate estimates of the time delays can be achieved. In this experiment, the residual noise can be expressed as:
\begin{equation}
\begin{aligned}
\gamma(\Lambda)=&{ \rm cor}_1(t_1)+{ \rm cor}_2(t_2)\\
=&{\rm7.667MHz}[q_2(t_1)\!-\!q_1(t_1)\\
&+\!q_1(t_1\!+\!\Lambda\!+\!\alpha t_1)\!-\!q_2(t_1\!+\!\Lambda\!+\!\alpha t_1)]. 
\end{aligned}
\end{equation}
Subsequently, the constant initial time offset is established by finding the minimum value of $\gamma(\Lambda)$ across various translation times $\Lambda$. The results are shown in Fig. \ref{fig:11}(b).

\section*{Declaration of Competing Interest}
The authors declare that they have no known competing financial 
interests or personal relationships that could have appeared to influence 
the work reported in this paper.

\section*{Acknowledgments}
This work is supported by National Key Research and Development Program of China (Grants No. 2023YFC2206100 and No. 2022YFC2204602), National Natural Science Foundation of China (Grant No. 12175076), and Knowledge Innovation Program of Wuhan-Basi Research (2023010201010048).

\section*{Data availability}
 Data underlying the results presented in this paper are not publicly available at this time but may be obtained from the authors upon reasonable request.



  \bibliographystyle{elsarticle-num} 
  \bibliography{elsarticle-template-num}
\end{document}